**PERSPECTIVE**  Open Access

# Perspective: imaging atomic step geometry to determine surface terminations of kagome materials and beyond

Guowei Liu[1†], Tianyu Yang[1†], Yu-Xiao Jiang[2†], Shafayat Hossain[2], Hanbin Deng[1*], M. Zahid Hasan[2,3,4,5*] and Jia-Xin Yin[1,6*]

**Abstract**

Here we review scanning tunneling microscopy research on the surface determination for various types of kagome materials, including 11-type (CoSn, FeSn, FeGe), 32-type ($Fe_3Sn_2$), 13-type ($Mn_3Sn$), 135-type ($AV_3Sb_5$, A = K, Rb, Cs), 166-type ($TbMn_6Sn_6$, $YMn_6Sn_6$ and $ScV_6Sn_6$), and 322-type ($Co_3Sn_2S_2$ and $Ni_3In_2Se_2$). We first demonstrate that the measured step height between different surfaces typically deviates from the expected value of ±0.4 ∼0.8Å, which is owing to the tunneling convolution effect with electronic states and becomes a serious issue for $Co_3Sn_2S_2$ where the expected Sn-S interlayer distance is 0.6Å. Hence, we put forward a general methodology for surface determination as atomic step geometry imaging, which is fundamental but also experimentally challenging to locate the step and to image with atomic precision. We discuss how this method can be used to resolve the surface termination puzzle in $Co_3Sn_2S_2$. This method provides a natural explanation for the existence of adatoms and vacancies, and beyond using unknown impurity states, we propose and use designer layer-selective substitutional chemical markers to confirm the validity of this method. Finally, we apply this method to determine the surface of a new kagome material $Ni_3In_2Se_2$, as a cousin of $Co_3Sn_2S_2$, and we image the underlying kagome geometry on the determined Se surface above the kagome layer, which directly visualizes the *p-d* hybridization physics. We emphasize that this general method does not rely on theory, but the determined surface identity can provide guidelines for first-principles calculations with adjustable parameters on the surface-dependent local density of states and quasi-particle interference patterns.

## 1 Introduction

Layered quantum materials are commonly cleaved into a variety of surface terminations. Determining these cleavage surfaces is fundamental because each surface carries its exclusive intrinsic properties to exhibit novel physical manifestations. The kagome material contains a kagome atomic layer with shared triangular lattices and harbors many novel physical phenomena, such as Dirac points, van Hove singularities, and flat bands [1–5]. In recent years, exotic phenomena such as superconductivity, spin density waves, and chiral electronic states emerging in kagome materials have attracted widespread attention in the condensed matter physics community [6–21]. Most intricate physical phenomena originate from the kagome layer [22, 23]. Therefore, determining the atomic surface is crucial for analyzing the electronic states and their coupling [24]. Scanning tunneling microscopy/spectroscopy (STM/S) can obtain local morphological structure and

*Correspondence: denghb@sustech.edu.cn; mzhasan@princeton.edu; yinjx@sustech.edu.cn
[1]Department of Physics, Southern University of Science and Technology, Shenzhen, China
[2]Department of Physics, Princeton University, Princeton, NJ, USA
Full list of author information is available at the end of the article †Equal contributors electronic state at the atomic level [25–27], which have played essential roles in exploring novel materials through their cleavage surfaces, such as A$E$Fe$_2$As$_2$ (A$E$ = $K$, Ca, Ba, La) [28–31], (Ba, Sr) Ni$_2$As$_2$ [32], URu$_2$Si$_2$ [33, 34], GdRu$_2$Si$_2$ [35], YbRh$_2$Si$_2$ [36–38], and CeCoIn$_5$ [39–41]. STM/STS is also one of the most effective research methods in studying kagome materials, especially, in determining their cleavage surfaces.

In most single-crystal STM research, a chunk is usually cryogenically cleaved in an ultra-high vacuum to obtain clean sample surfaces. To identify the surfaces, the straightforward methods include measuring the height difference at atomic steps [42] or atomically identifying the symmetry and lattice constants, and then comparing them with the lattice structure, which proves the fundamental evidence for the determination of the cleavage surfaces and provides an essential prerequisite for many experimental analyses, such as quasi-particle interference [23, 43]. However, the complex tunneling matrix element affects the distance between the tip and surface [44], making it difficult to determine the interlayer distance precisely (details see below). Furthermore, the surface reconstruction may change the lattice constant as well as the symmetry of the surface structure, making it inappropriate to correspond with the bulk lattice structure. These reasons make the determination of cleavage surfaces in some kagome materials challenging [23, 24, 30, 33, 34, 45–55]. Therefore, the academic community urgently needs an experimental method with high versatility and accuracy to determine the cleavage surfaces of kagome materials through STM/STS.

## 2 Imaging atomic step geometry

Here, we put forward a plain vanilla and general method to determine the cleavage surfaces by comprehensively imaging kagome materials' atomic step geometry. The recently studied kagome materials are mostly layered structures, and we classify their atomic layers into three types: kagome layers ($K$), neighboring layers ($N$), and isolated layers ($I$) (Fig. 1a). Atomic kagome ($K$) layers usually comprise smaller atoms, with dense atomic arrangement, high state density, and tight chemical bonds. Neighboring layers ($N$) are next to kagome layers in the crystal and are bonded to kagome layers or have strong interlayer interactions. Isolated layers ($I$) are relatively far from the kagome layers in the crystal and have weak interlayer interactions with adjacent atomic layers. Their intervals indicate the interlayer bonding strength. Since $K$ layers have tight bonds, and the stronger bonding of the $N$-$K$-$N$ trilayered block, the cleavage almost occurs between the neighboring layers $N$ and isolated layer $I$ by breaking the interlayer and intralayer bonds of $I$, which yields an atomic step with vacancy defects and adatom defects on both sides simultaneously and symmetrically (Fig. 1b).

Based on the simple cleavage scheme of kagome materials (Fig. 1a-1b), the atomic step edge offers abundant facts to explore, such as step height, defects, interlayer relationship, edge state, and so on. The step height usually cannot give a precise reference, since the discrepancy of electronic states between two adjacent layers in kagome materials always exists, and the tunneling measurement convolutes the local density of states. As shown in Fig. 1c (and d), since the electronic state of the lower $N$ layer is higher (lower) than the upper $I$ layer, the measured $h$ is lower (higher) than the expected layer distance $d$. Surprisingly, we discover that the defect types are always helpful in determining the cleavage surfaces due to the interlayer relationship. From the cleavage scheme, vacancy defects always appear on the bonds broken layer ($I$) (Fig. 1e), meanwhile, the adatom defect always remains on the neighboring layer ($N$) (Fig. 1f). In fact, the defect types provide evidence to determine cleavage surfaces to some extent. Another independent piece of evidence is from the designer layer selective substitutional chemical maker. As shown in Fig. 1g, the atoms of the selected layer can be replaced equivalently by the same group of element atoms, thus, identifying the cleavage surfaces. This paper put forward a universal atomic step geometry imaging methods to determine the cleavage surfaces by reviewing previous reports, including 11-type (CoSn), 32-type (Fe$_3$Sn$_2$), 13-type (Mn$_3$Sn) [56–58], 135-type (KV$_3$Sb$_5$), 166-type (TbMn$_6$Sn$_6$), and 322-type (Co$_3$Sn$_2$S$_2$) [22, 30, 47, 49, 52–54, 59–62], and the latest STM research results on 166-type (ScV$_6$Sn$_6$), and 322-type (Ni$_3$In$_2$Se$_2$). We also note that the atomic step geometry imaging method sounds fundamental but it is also challenging to implement, requiring patience for scanning large areas and even several samples to locate the atomic step (that can take several weeks) and exceptional experimental skills for precise imaging with atomic resolution and without muti-tip effects. It is our experience that with many adjustable parameters, first principles can be consistent with both cases when the experiment results on surface terminations are controversial. Therefore, it is crucial to first determine the surface identity in a self-consistent experimental way without referring to theory, after which, the determined surface identity can be used as a guideline for further first-principles calculations (or other model analysis) on surface-dependent density of states and quasi-particle interferences.

## 3 Surface determination of multiple layered kagome materials

### 3.1 11-type

The 11-type including CoSn [50, 60, 63–65], FeSn [53, 54, 66, 67], and FeGe [19, 59, 62, 68–71], only contains one kagome layer and one honeycomb layer. We start from the simplest bilayer kagome lattice CoSn (space group P6/mmm). It consists of a Co$_3$Sn kagome layer (Fig. 2a left panel, $K$ layer) and a Sn$_2$ honeycomb layer (Fig. 2a right panel, $I$ layer) with alternating stacking and no $N$

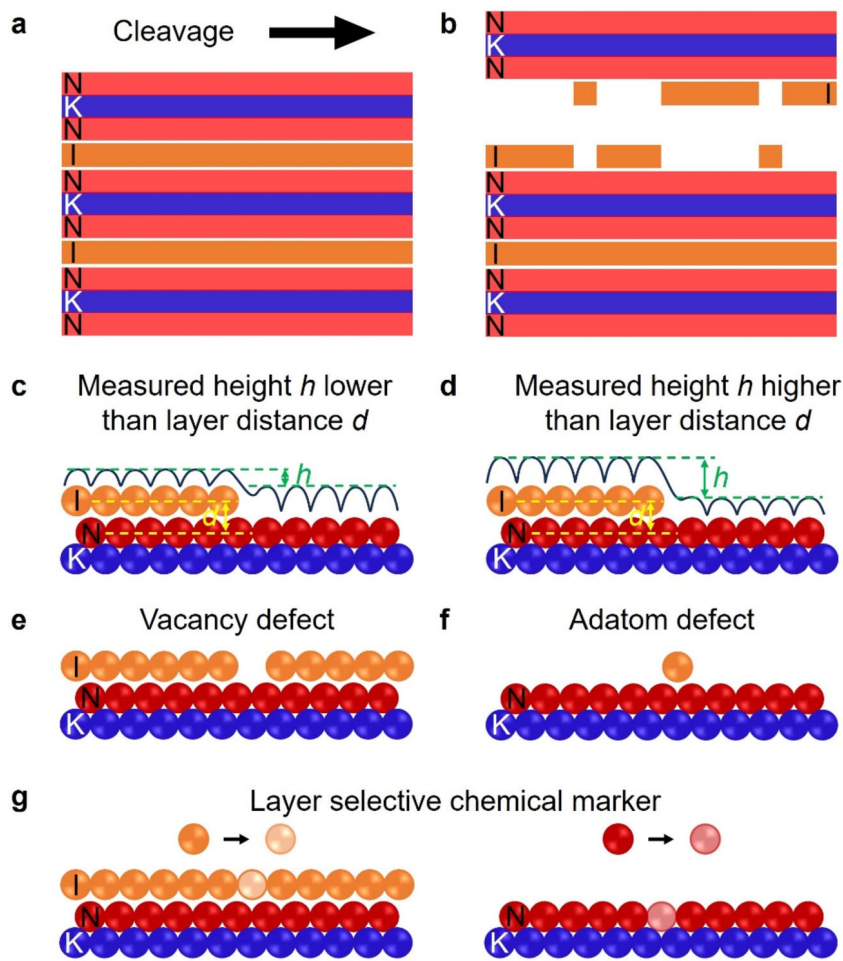

**Figure 1** *Schematics and identifications of atomic step cleavage.* **a** A typical kagome model is perioded with a trilayered strong-bonding *N-K-N* block further sandwiched by the isolated weak-bonding *I* layers. **b** Schematic diagram illustrating a typical cleavage process of a kagome bulk. Cleavage occurs between the *N* layers and *I* layer, yielding vacancy defects on the *I* layer and adatom defects on the *N* layer at both cleaving sides simultaneously and symmetrically. **c, d** Unreliable height determination method. The measured *h* might be lower than the expected layer distance *d*, since the lower layer has a higher density of state than that of the upper layer **c**; or reverse **d**. **e, f** Identification of cleavage surfaces by different types of defects. The upper broken *I* layer produces vacancy defect **e**, and remains adatom defect on the lower *N* layer **f**. **g** Identification of cleavage surfaces by layer selective chemical marker. The atoms of the selective layer are substituted by the dopants from the same group element

layer in between. Consequently, cryogenically cleaving a CoSn single crystal just yields $Co_3Sn$ kagome or $Sn_2$ honeycomb surfaces. Utilizing atomic-level ultra-high resolution STM imaging, the $Co_3Sn$ surface with kagome feature (Fig. 2b left panel) and the $Sn_2$ surface with honeycomb feature (Fig. 2b right panel) have been observed unambiguously [50, 60], respectively. It is also a rare case that all the Co kagome atoms and central Sn atoms are resolved individually. In most cases, the imaging of $TM_3Sn$ (TM is the transition metal) layer yields a hexagonal symmetry lattice with Sn atomic position as a dark spot. In fact, if one performs a direct first-principles simulation for the $TM_3Sn$ topographic image, one will often find that the highest spot should be the Sn atom position as the Sn atom is larger. This discrepancy highlights the correlated nature of these kagome materials that often go beyond the simple first-principles simulation, and warns us of the importance of experimentally determining the atomic details before seeking help from first-principles. If the conclusion cannot be drawn from the experimental data independently, then it can often lead to systematic errors with combined scanning tunneling microscopy experiments and first-principles calculations, which is at least true in our research experience.

As shown in Fig. 2c, a step schematic is generated by breaking the bonds between the $Co_3Sn$ and the $Sn_2$ layers as well as the bonds in the $Sn_2$ intralayer, where the expected step height is half of the *c*-axis lattice constant.

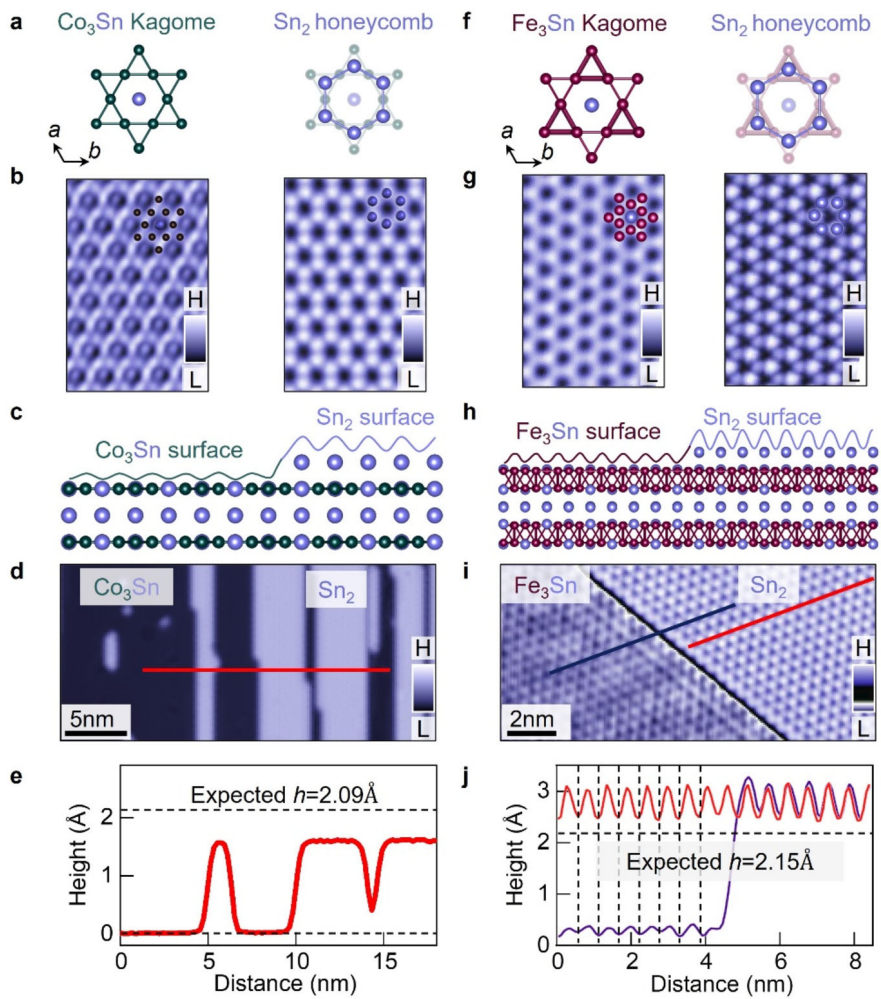

**Figure 2** *Atomic-scale surface identification of CoSn and Fe₃Sn₂*. *a* Kagome lattice structure of the Co$_3$Sn layer (left panel) and the adjacent honeycomb Sn$_2$ layer (right panel). *b* Atomically resolved topographic image with the corresponding atomic lattice structures of the Co$_3$Sn surface (left panel) and Sn$_2$ surface (right panel) (5 × 7.5 nm$^2$, $V = -10$ mV, $I = 8$ nA). *c* Side view of the crystal structure and illustration of the two possible terminating surfaces. The curve illustrates the surface profile. *d* Topographic image of a single atomic step between two surfaces (30 × 15 nm$^2$, $V = 100$ mV, $I = 0.3$ nA). *e* The topological profile cutting the step along the line marked in *d* shows a step height of ∼1.6 Å, while the expected step height is 2.09 Å. Adapted from [60]. *f* Kagome lattice structure of the Fe$_3$Sn layer (left panel) and the adjacent honeycomb Sn$_2$ layers (right panel). *g* Atomically resolved Fe$_3$Sn (left panel) and Sn$_2$ (right panel) surfaces. The insets show the corresponding atomic lattice structures (5 × 7.5 nm$^2$, $V = 50$ mV, $I = 0.8$ nA). *h* Side view of the crystal structure and illustration of the two possible terminating surfaces. The curve illustrates the surface profile. *i* Lattice alignment between the Sn$_2$ surface and the Fe$_3$Sn surface from an edge of a Sn surface step (15 × 7.5 nm$^2$, $V = 50$ mV, $I = 0.8$ nA). *j* Comparison of the line-cut profiles taken along the red and blue lines marked in *i*. Adapted from [52]

Figure 2d presents a representative single atomic step between the lower Co$_3$Sn kagome layer and the upper Sn$_2$ honeycomb layer. The step height is measured as 1.6 Å smaller than the expected height of 2.09 Å [63] in Fig. 2e. The discrepancy comes from the higher integrated density of states of the Co$_3$Sn kagome layer than the Sn$_2$ honeycomb layer at the bias voltage of 100 mV [60].

### 3.2 32-type
Correspondingly, Fe$_3$Sn$_2$ (space group $R\bar{3}m$) is another kind of trilayer kagome lattice structure without *N* layer, which only consists of a Fe$_3$Sn kagome bilayer (Fig. 2f left panel, *K* layer) and a Sn$_2$ honeycomb layer (Fig. 2f right panel, *I* layer) in a unit cell. In our high-resolution STM images, we can distinguish the Fe$_3$Sn kagome surface (Fig. 2g left panel) from the Sn$_2$ honeycomb surface (Fig. 2g right panel). Similar to CoSn, a dilemma of step height appears in the Fe$_3$Sn$_2$ system. Due to the density of states of the Sn$_2$ honeycomb layer being higher than that of the Fe$_3$Sn kagome layer at a bias voltage of 50 mV [52], the step height of the Sn$_2$ layer on the Fe$_3$Sn layer measured as 3.0 Å, which is larger than the expected value of 2.15 Å.

The constant current topographic image convolutes the spatial variation of the integrated local density of states (LDOS) and the geometrical corrugations, the step height measured by the STM tip cannot become conclusive evidence for surface determinations [30].

In the crystal structure, the $Sn_2$ honeycomb layer has perfect $C_6$ symmetry, while the $Fe_3Sn$ kagome bilayer is $C_3$ symmetry. It seems that the two surfaces can also be distinguished by lattice symmetry. However, due to electronic perturbation of the underneath breath kagome bilayer, the perfect $C_6$ symmetry of the $Sn_2$ honeycomb lattice is also reduced into the $C_3$ symmetry in the STM image. Therefore, in STM images with low resolution, the $Sn_2$ honeycomb lattice may be misidentified with $Fe_3Sn$ kagome bilayer. While in our high-resolution STM image, the atoms in the $Sn_2$ honeycomb termination are all resolved, with alternative brightness and dimness feature points (Fig. 2g right panel). The ultra-high-resolution STM image of the atomic step further determines lattice alignment (Figs. 2i and j), showing that the dark position of the $Fe_3Sn$ layer aligns with the dark position of the $Sn_2$ layer. This allows us to assign the Fe and Sn atoms for the topographic image of $Fe_3Sn$ layer in the left panel of Fig. 2g. Again, here the dark position in the $Fe_3Sn$ layer corresponds to the Sn atom, and the first-principles calculation can often assign the Sn atom to the highest position in the topographic image of $Fe_3Sn$ layer, opposite to the experimentally determined case.

### 3.3 135-type

For the above two simple cases, high-resolution imaging of atomic step provides a feasible method to identify cleavage surfaces, rather than step height. The same approach applies to the well-known 135-type [72], for instance, $CsV_3Sb_5$ [17, 73–75], $KV_3Sb_5$ [7, 18, 61, 76, 77], $RbV_3Sb_5$ [8, 78–80], $Cs(V, Ta)_3Sb_5$ [21, 81]. As Fig. 3a shows, a typical crystal structure of $KV_3Sb_5$ consists of a vanadium-based kagome layer (*K* layer) and two sandwiched antimonene (Sb) honeycomb lattices (*N* layer). The *I* layers are hexagonal alkali metal lattices with the largest intralayer bond length, making them fragile for cleavage. According to the crystalline symmetry, the cleavage between the alkali metal layer and the $Sb_2$ layer yields a step with the upper alkali layer and the lower Sb layer (Fig. 3b). Figure 3c shows a typical gradient atomic step edge, where the Sb surface is transiting to the K surface from left to right. Atomically resolved topographies exhibit a close-knit honeycomb lattice of lower Sb surface (Fig. 3d) and hexagonal lattice of upper K surface [78, 82] (Fig. 3e), respectively, in which a 2 × 2 charge order wave modulates both layers and breaks the translational symmetry. We also note that the weakly bonded alkali islands or adatoms in 135-type of materials can be swept away by the STM tip when the tip is very close to the surface, which can help to create a large area of Sb surface for further spectroscopic research [73].

### 3.4 166-type

Another kagome lattice worth discussing in detail is the 166-type for its different interlayer structure [83–87]. Figure 4a shows a unit lattice structure of $ScV_6Sn_6$, a typical 166-kagome lattice, in which the blue kagome layers are made of vanadium (V) atoms. Three contiguous Sn layers form triangular ($Sn_2$ surface), honeycomb ($Sn_1$ surface), and triangular ($Sn_2$ surface) structures, respectively. Between V layers, a sharing Sc-Sn mix-layer has large interlayer distances with these two kagome layers symmetrically, thus, a cleave can occur between them in this case. Considering our previous points, in this system, cleavage can occur between any layers, forming all four terminations. Here, we found three consecutive atomic steps of $ScV_6Sn_6$ [83]. From low to high in Fig. 4c, they are honeycomb, triangle, and tripped structures, corresponding to the $Sn_1$ layers, $Sn_2$ layers, and V layers respectively. Their magnifying images are shown in Fig. 4d-f. Due to the small size of V atoms, the resolution of the kagome lattice in this system is insufficient to display, so we have presented the Mn kagome lattice layer of $TbMn_6Sn_6$ with a larger atomic size, and its high-resolution kagome lattice is shown in the Fig. 4g [22]. Additionally, our previously published article showed a possible Tb-Sn layer of $TbMn_6Sn_6$ that corresponds to the Sc-Sn layer in $ScV_6Sn_6$ not present here [22].

### 3.5 322-type

We have discussed the cleavage surfaces of different stacking kagome lattices from honeycomb-kagome-honeycomb (11-type), honeycomb-kagome-kagome-honeycomb (32-type), hexagon-honeycomb-kagome-honeycomb-hexagon (135-type) to super-complex-interlayer (166-type). There is a consensus in these systems, making them a good starting point for the atomic step geometry imaging method. Now we take a step forward to identify the controversial cleavage surface of another layered kagome lattice $Co_3Sn_2S_2$ (Fig. 5a) [23, 24, 45–49]. $Co_3Sn_2S_2$ (space group $R\bar{3}m$) has similar stacking behavior to 135-type, except for hexagon-hexagon-kagome-hexagon-hexagon, which manifests Sn- and S- terminated non-kagome surfaces with the identical lattice constant and hexagonal symmetry. In this case, a single atomic step is indispensable for identifying these surfaces.

By summarizing the three above cases, the cleavage always occurs at the interval between the *N* (or *K*) layer and *I* layer, due to the weaker chemical bonding of the *N* (or *K*)-*I* interlayer and *I* intralayer. Hence, in the step edge, the *I* layer forms vacancy defects, such as $Sn_2$ and alkali metal surfaces, while the *N* (or *K*) layer forms adatom defects, such as $Co_3Sn$ kagome, $Fe_3Sn$ kagome, and $Sb_2$ honeycomb surfaces. Applying this experience to $Co_3Sn_2S_2$,

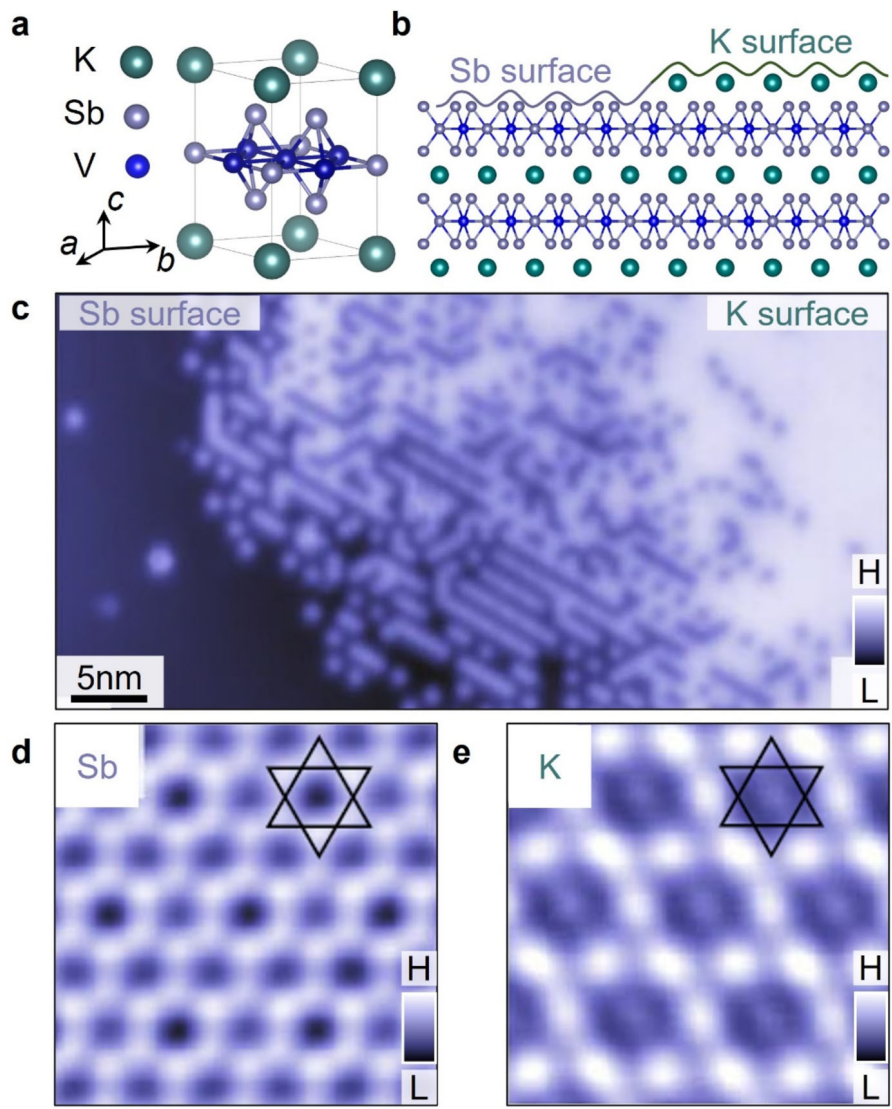

**Figure 3** *Surface identification of KV$_3$Sb$_5$ at the atomic scale. a* Crystal structure of KV$_3$Sb$_5$. *b* Cleavage surfaces of KV$_3$Sb$_5$ illustrated from the side view of the crystal structure. *c* Topographic image of a surface step edge, containing both K (right region) and Sb (left region) surfaces (54 × 27 nm$^2$, $V$ = 100 mV, $I$ = 0.05 nA). *d* Atomically resolved topographic images of the Sb honeycomb surface (2.8 × 2.8 nm$^2$, $V$ = 100 mV, $I$ = 0.5 nA). *e* Atomically resolved topographic images of the K hexagonal surface (2.8 × 2.8 nm$^2$, $V$ = 100 mV, $I$ = 0.5 nA). The black lines marked in *d* and *e* denote the underlying kagome lattice. Adapted from [61]

it will be brief to determine the cleavage surfaces when encountering a promising atomic step. As illustrated in Fig. 5b, the bonds between the Sn neighboring layer and the S isolated layer are broken. Similar to the three previous cleavage cases, the upper $I$ layer forms vacancies, and the lower $N$ layer forms adatoms. The high-resolution STM image of an ideal step edge of Co$_3$Sn$_2$S$_2$ shown in Fig. 5c demonstrates the upper Sn isolated surface with vacancy defects and the lower S neighboring surface with adatom defects. Moreover, Sn has smaller corrugation than the S surface, due to the atomic radius of Sn being larger than S, which further proves the surface identification from the high-resolution images of the S surface and the Sn surface in Figs. 5d-5e, respectively. We have also been aware that Ref. [23] mainly uses the step height argument for the surface identification, and it is clear from Fig. 2e, g and the basic principle of STM that the topographic height convolutes the LDOS and the geometrical corrugations that the step height argument is not decisive for termination assignment.

To further confirm the assignation, a chemical-marker experiment is designed such that 1% In impurities are doped into $Co_3Sn_2S_2$ bulk [39], where In dopants equivalently substituted Sn preferentially [88, 89]. Figure 6a demonstrates a high-resolution topographic image showing that the layer selective In-dopants with consistent concentration are distributed sparsely on the Sn-terminated surface. Moreover, from the high-resolution image of the typical defect (Fig. 6b), the substituted site is consistent with the topmost Sn site without lattice mismatch, which proves this surface with vacancy defects is the Sn surface. Following Ref. [23], the imaged impurity states shape with substantial multi-tip effects from unknown underlying (not at the surface) defects are used to provide hints for surface terminations [43]. The chemical identity, atomic position, and tunneling matrix element of the underlying defects are all unknown. Even for known underlying impurities such as Zn in cuprates [90], it is well known that the simple model would not produce the correct symmetry of the impurity states and the tunneling matrix element has to be considered. Therefore, its accuracy in identifying surface terminations with unknown underlying defects is not at the same level as the chemical marker experiment [30, 39]. We further propose using 1% Se to replace S in $Co_3Sn_2S_2$ to mark the S layer as a complementary layer-selective chemical marker experiment.

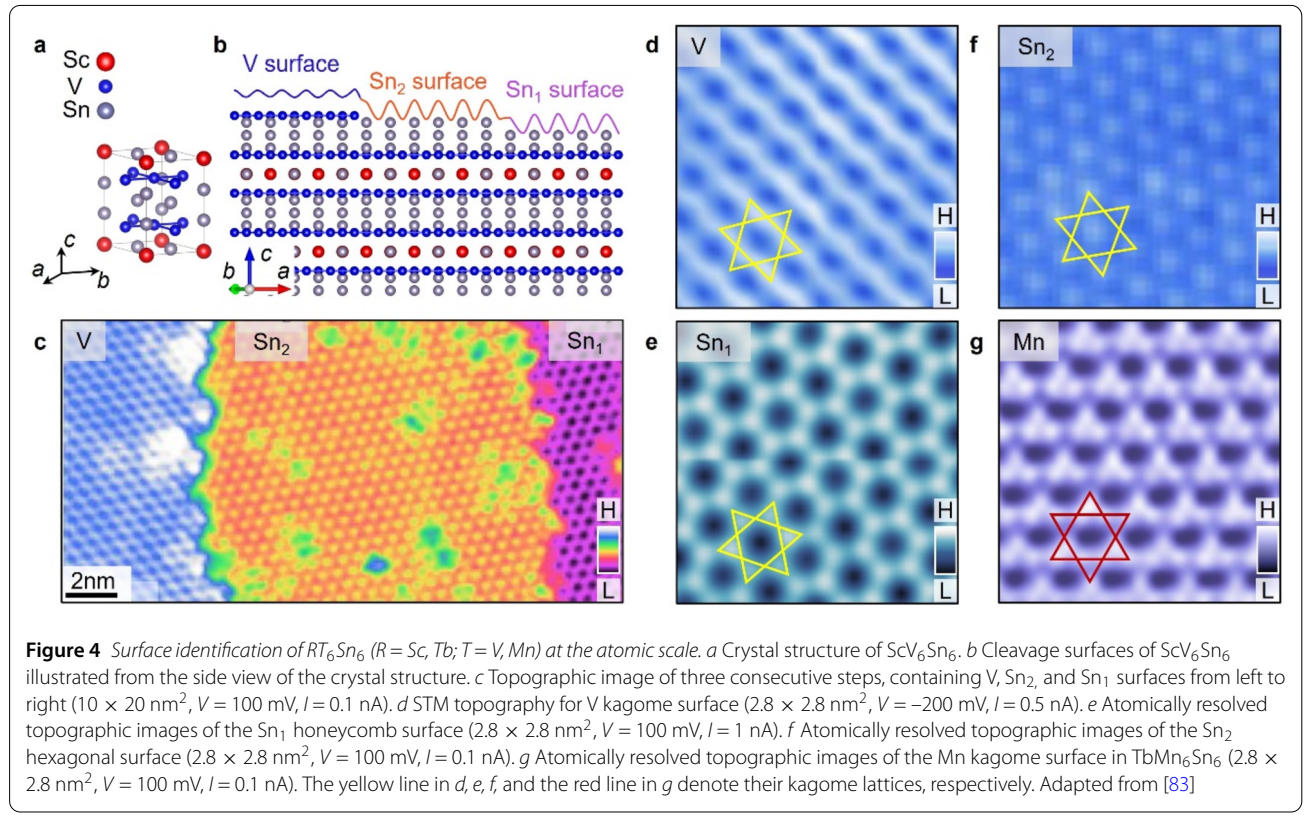

**Figure 4** *Surface identification of $RT_6Sn_6$ (R = Sc, Tb; T = V, Mn) at the atomic scale. a Crystal structure of $ScV_6Sn_6$. b Cleavage surfaces of $ScV_6Sn_6$ illustrated from the side view of the crystal structure. c Topographic image of three consecutive steps, containing V, $Sn_2$, and $Sn_1$ surfaces from left to right (10 × 20 $nm^2$, V = 100 mV, I = 0.1 nA). d STM topography for V kagome surface (2.8 × 2.8 $nm^2$, V = −200 mV, I = 0.5 nA). e Atomically resolved topographic images of the $Sn_1$ honeycomb surface (2.8 × 2.8 $nm^2$, V = 100 mV, I = 1 nA). f Atomically resolved topographic images of the $Sn_2$ hexagonal surface (2.8 × 2.8 $nm^2$, V = 100 mV, I = 0.1 nA). g Atomically resolved topographic images of the Mn kagome surface in $TbMn_6Sn_6$ (2.8 × 2.8 $nm^2$, V = 100 mV, I = 0.1 nA). The yellow line in d, e, f, and the red line in g denote their kagome lattices, respectively. Adapted from* [83]

Recently, the qPlus non-contact atomic force microscopy [24] has also revealed differences between the two terminations. However, the experimental data would probably not be capable of assigning the surface identifications directly and independently. Unlike the step edge imaging method, it has not been put as a general for determining surfaces in other kagome materials and other quantum materials. The first-principles calculations are then used for guiding the surface identification by matching with the atomic force microscopy data. It was our experience in 122 iron-based superconductors [30] that depending on the adjusting parameters, first principles can assign the surface terminations quite differently and oppositely. Its reliability in identifying the surfaces 122 iron-based superconductors is in doubt in the community. Even in the case of $Co_3Sn_2S_2$, first-principles calculated LDOS can always match with the experimentally observed ones with both scenarios depending on the choice of shifting the Fermi energy within ±500 meV, substantially reducing its credibility. It is our understanding that in the electronic structure community including angle-resolved photoemission and scanning tunneling microscopy, researchers including ourselves often seek help from first principles when the experimental data alone is inconclusive and our subjective interpretation of the data can guide the parameter tuning directions of the first-principles. Therefore, a first-principles independent methodology as demonstrated here is very crucial to firstly set up the correct and objective directions for the adjustable parameter tuning in the first-principles. Although STM cannot directly re-

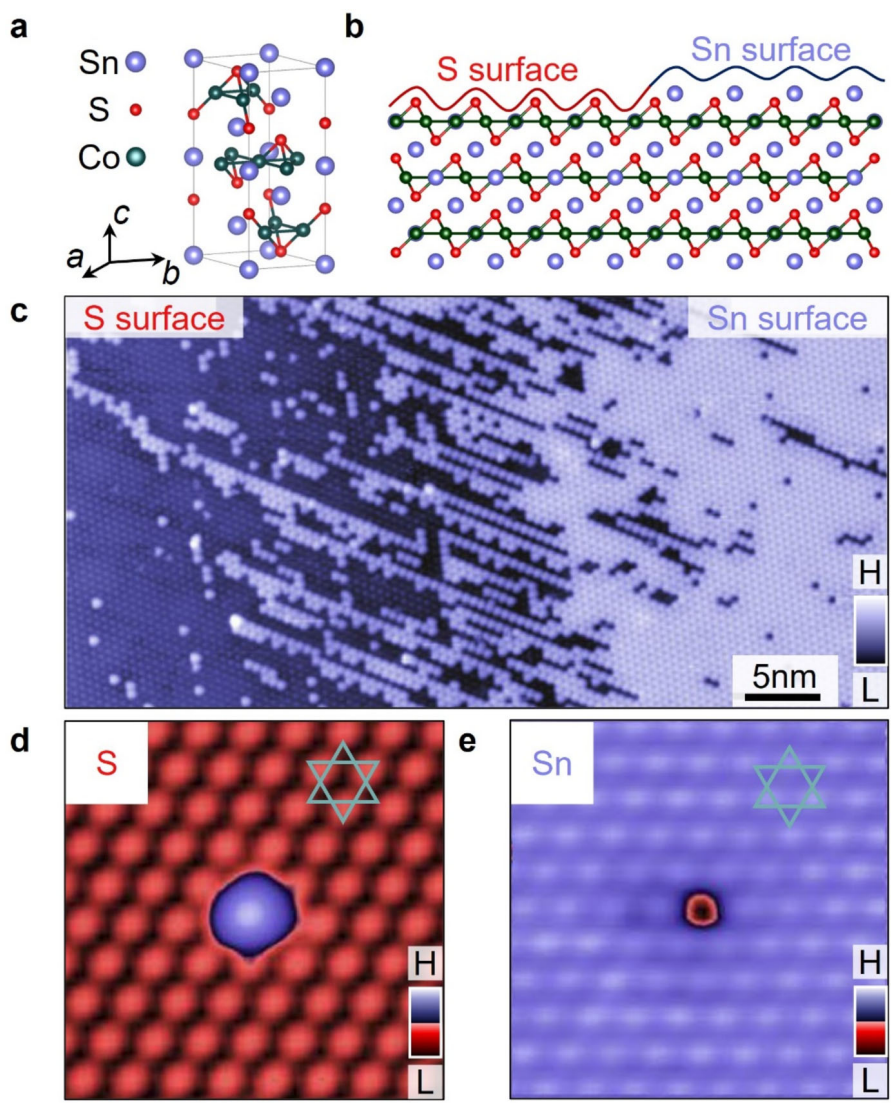

**Figure 5** *Atomic-scale visualization of the cleavage surface of $Co_3Sn_2S_2$. a* Crystal structure of $Co_3Sn_2S_2$. *b* Cleavage surfaces of $Co_3Sn_2S_2$ illustrated from the side view of the crystal structure. *c* Atomically resolved topographic image of the boundary between the S surface and Sn surface (54 × 27 nm$^2$). The S surface smoothly evolves into the Sn surface with increasing coverage of the Sn adatom. *d* The common defect on the S surface is Sn adatom (4 × 4 nm$^2$). *e* The common defect on the Sn surface is Sn vacancy (4 × 4 nm$^2$). (all images recorded at $V = -50$ mV, $I = 0.2$ nA). Adapted from [47]

solve the surface atomic element, it can detect the electronic structure beneath the surface through tunneling, which provides a chance to extract the information from the underlying kagome layer. Therefore, we chose another 322-type analogue, nominating $Ni_3In_2Se_2$ [91], which has the same lattice structure as $Co_3Sn_2S_2$. Figure 7a shows the side view of the step schematic model of $Ni_3In_2Se_2$, where the cleavage occurs between the In isolated layer and the Se neighboring layer. Figure 7b shows the typical step of the In and Se surface. Similar to $Co_3Sn_2S_2$, the left upper isolated layer is defined as In surface with typical vacancy defect, meanwhile, the right lower neighbor-ing layer is defined as Se surface with In adatom defects. Although they have identical lattice symmetry, their electronic states affected by the underlying kagome layer are different. In surface is far away from the $Ni_3In$ kagome layer, so its STM topographic images show a stubborn triangular lattice for both positive and negative bias voltage (Fig. 7c), whereas, the lower Se surface directly bonding to the $Ni_3In$ kagome layer manifests a kagome lattice feature (Fig. 7d). Hence, we observed an apparent kagome symmetry on the Se surface at the bias voltage of −100 mV and −200 mV (Fig. 7d), whereas, an affected triangular lattice is shown in the atomic resolution image at a bias voltage

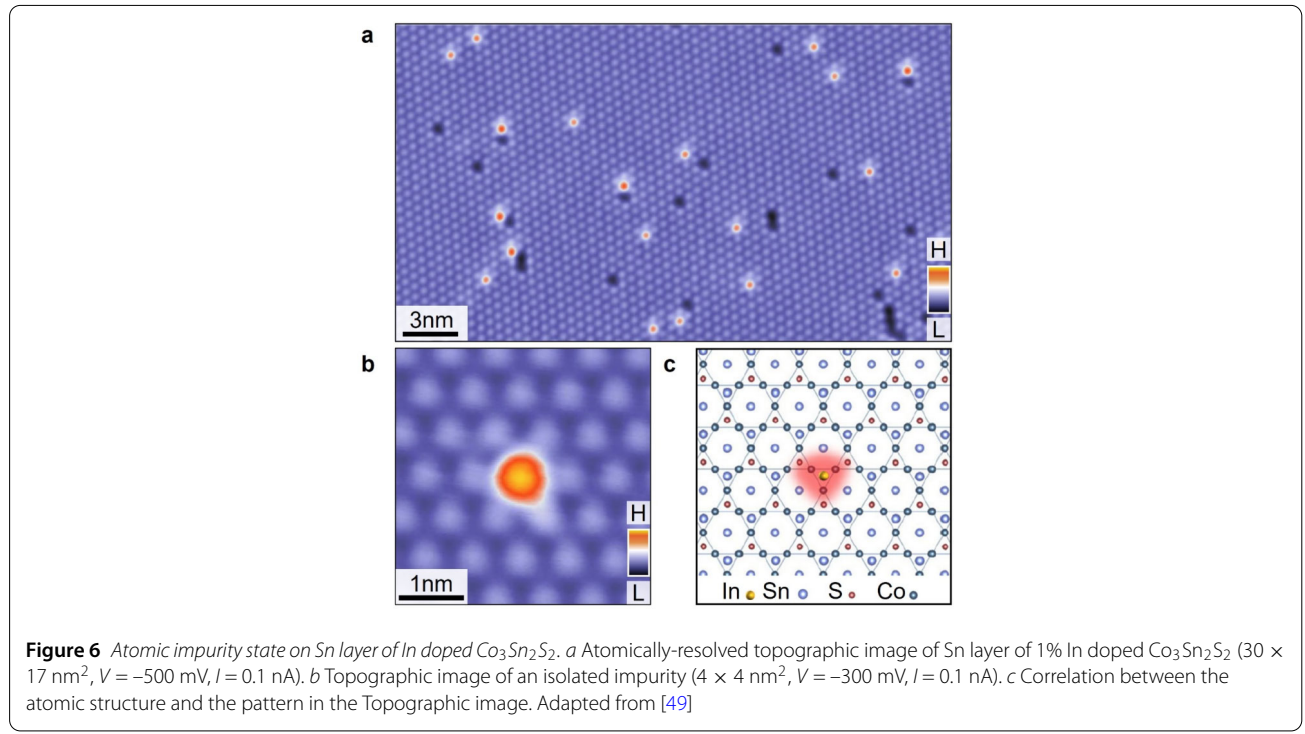

**Figure 6** *Atomic impurity state on Sn layer of In doped $Co_3Sn_2S_2$*. *a* Atomically-resolved topographic image of Sn layer of 1% In doped $Co_3Sn_2S_2$ (30 × 17 $nm^2$, $V$ = –500 mV, $I$ = 0.1 nA). *b* Topographic image of an isolated impurity (4 × 4 $nm^2$, $V$ = –300 mV, $I$ = 0.1 nA). *c* Correlation between the atomic structure and the pattern in the Topographic image. Adapted from [49]

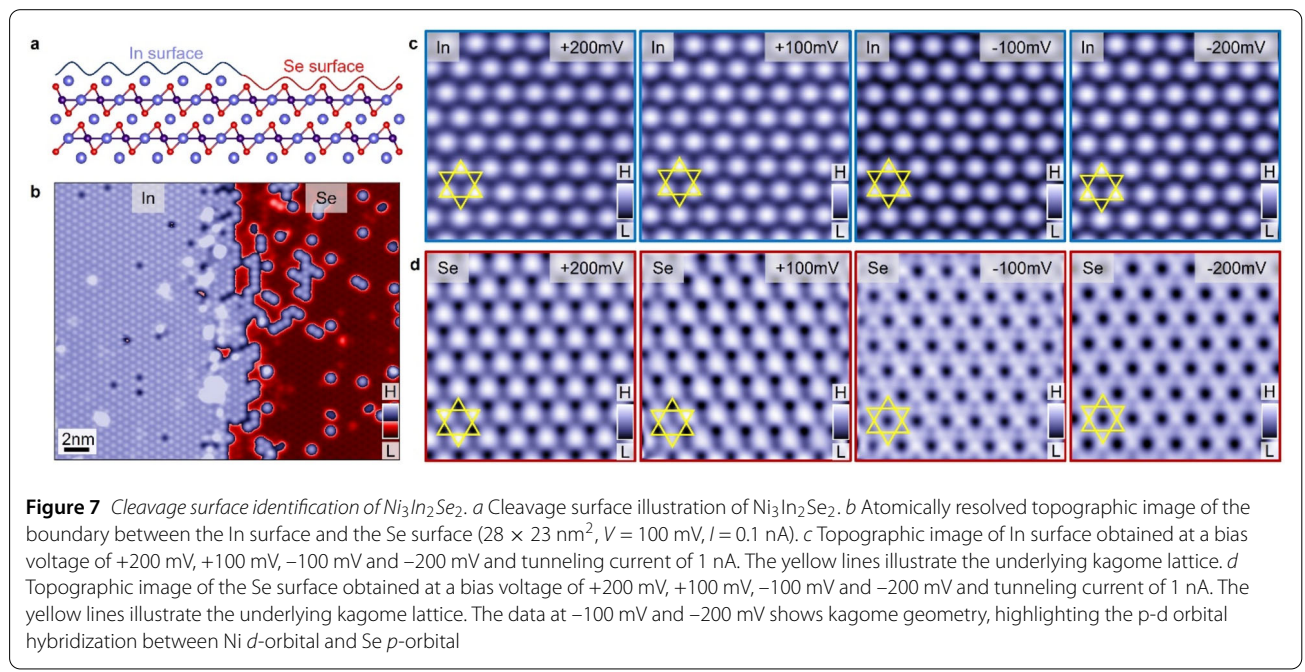

**Figure 7** *Cleavage surface identification of $Ni_3In_2Se_2$*. *a* Cleavage surface illustration of $Ni_3In_2Se_2$. *b* Atomically resolved topographic image of the boundary between the In surface and the Se surface (28 × 23 $nm^2$, $V$ = 100 mV, $I$ = 0.1 nA). *c* Topographic image of In surface obtained at a bias voltage of +200 mV, +100 mV, –100 mV and –200 mV and tunneling current of 1 nA. The yellow lines illustrate the underlying kagome lattice. *d* Topographic image of the Se surface obtained at a bias voltage of +200 mV, +100 mV, –100 mV and –200 mV and tunneling current of 1 nA. The yellow lines illustrate the underlying kagome lattice. The data at –100 mV and –200 mV shows kagome geometry, highlighting the p-d orbital hybridization between Ni *d*-orbital and Se *p*-orbital

of +100 mV and +200 mV (Fig. 7d). Similar to Si(111)-7 × 7, lower negative bias voltage imaging facilitates the visualization of the underlying kagome layer, because their electronic states act on tunneling at this time [92, 93]. Our observation of kagome geometry on the Se surface also directly visualized the *p-d* orbital hybridization physics as discussed in the atomic force microscopy work [24]. In short, this experimental approach and evidence provide another theory-independent way for surface identification in the 322 family.

### 3.6  31-type

We discuss the surface termination of kagome antiferromagnet $Mn_3Sn$ [56–58], which is well known for its gi-ant room temperature anomalous Hall effect and potential in antiferromagnetic spintronics application. The atomic Mn$_3$Sn kagome layers ($K$) directly stack with each other with AB stacking order (Figs. 8 a and b). Because the interlayer bond distance is similar to the in-plane bond distance, the cryogenic cleavage of the Mn$_3$Sn crystals would not yield a flat surface with atomic lattice resolution, but disordered surfaces. To prepare a large atomic flat surface, we can either anneal the surface *in-situ* with a high temperature (1100 K) for hours or apply a large pulse voltage (10 V) to bomb out a fresh surface [56, 58]. The topographic image of the Mn$_3$Sn layer (Fig. 8c) is very similar to the Fe$_3$Sn layer in Fe$_3$Sn$_2$. For each kagome pattern, it can be seen from the topographic image that the symmetry is reduced to C$_3$ from bulk C$_6$. This is likely due to the impact of the AB stacking, and the existence of the underlying Sn atom could play a role (lower panel of Fig. 8b) which can be studied by imaging atomic step geometry.

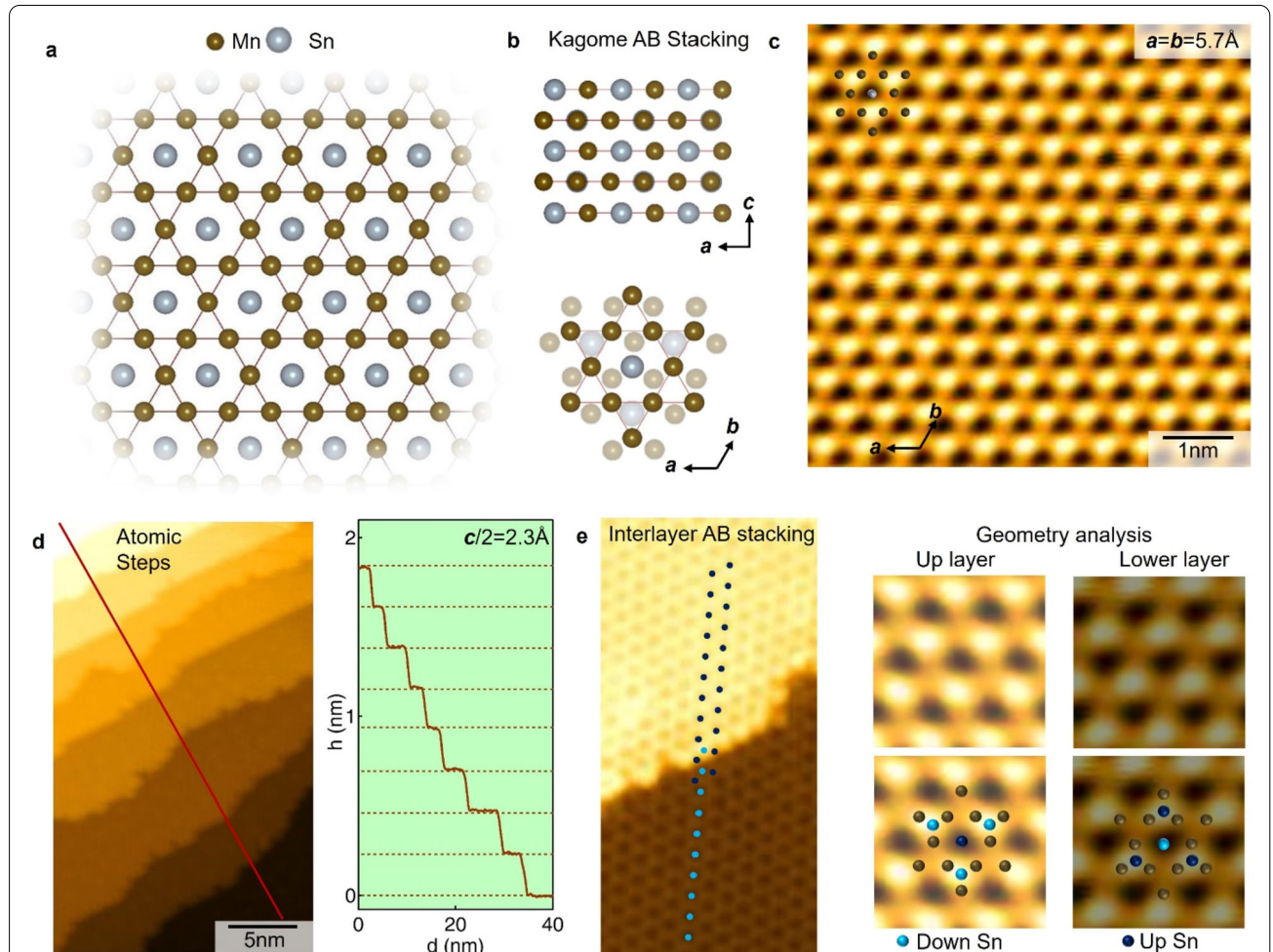

**Figure 8** *Atomic identification of Mn$_3$Sn.* a Crystal structure of the Mn$_3$Sn kagome lattice. b c-axis AB stacking of the kagome lattice from side view and top view, respectively. c STM topographic images of Mn$_3$Sn showing the lattice with hexagonal symmetry. d STM topographic image across multiple steps of single layers, whose line profile is shown on the right. e Left panel shows the stacking alignment of two layers across a step, with the black and blue dots denoting Sn atoms of the upper and lower layers, respectively. Right panel shows their atomically resolved topographic images with the corresponding atomic lattices. All topographic data were taken at T = 4.6 K, V = −50 mV, I = 50 pA. Adapted from [56]

Figure 8d shows a series of atomic steps. Because every step is of the Mn$_3$Sn layer, the convolution effect of the local density of states cancels out and the step height in this case matches the expected value. Then we focus on one atomic step in Fig. 8e. We first mark the position of the Sn atom for both layers corresponding to the periodic dark spots. It can be seen that the positions of the Sn atoms from the two layers do not overlap, supporting the AB stacking structure. Then we zoom into both layers to observe the C$_3$ kagome pattern and mark the position of the Sn atom from the sublayer. This geometry analysis based on experimental data demonstrates that the coupling with the underlying Sn atom makes the kagome pattern C$_3$ symmetric.

## 4  Conclusions

We have summarized the theory-independent atomic step geometry imaging methods to identify cleavage surfaces in various kagome lattices. High-resolution imaging of atomic steps on different terminations provides a straightforward judgment to determine the cleavage surfaces. Terminations with the same lattice symmetry and lattice constant are not so obvious to distinguish. Just like the atomic step height between two vicinal terraces convoluted with disparate electronic properties does not provide a precise value corresponding to the crystal structure. Both criteria are ambiguous in determining the cleavage surfaces of kagome materials. To settle this problem, we did a time-consuming scouring of the atomic steps of various kagome lattices. First, we have obtained the lattice alignments through the across-step profiles of height. Secondly, we have simultaneously recorded the topographies of the upper and the lower surfaces at different bias voltages. Thirdly, the continuous step shows the formation and evolution of defects on the upper and the lower surfaces. From $CoSn$, $Fe_3Sn_2$, $KV_3Sb_5$ $ScV_6Sn_6$ and $TbMn_6Sn_6$, to $Co_3Sn_2S_2$ and $Ni_3In_2Se_2$, all results show that the surfaces with vacancy defects are the In surface or Sn surface, and the surfaces with adatom defects are the S surface or Se surface in 322-type. The methodology provided here is general and consistent for different kagome cleavages and will be beneficial to the rapid development of new kagome materials and other layered quantum materials.

### Acknowledgements
We acknowledge that Shuheng Pan and Ang Li for proposing this method in studying surface terminations of iron-based superconductors. This perspective is written to celebrate Shuheng Pan's 75 birthday.

### Author contributions
GL, TY and JXY wrote the manuscript with contributions from all authors. GL, TY and HD performed STM measurements on $Ni_2In_2Se_2$. YJ and MSH performed STM measurements on $ScV_6Sn_6$. HD, MZH, and JXY supervised the project. All authors read and approved the final manuscript.

### Funding
J.X.Y. acknowledges the support from the National Key R&D Program of China (No. 2023YFA1407300) and the National Science Foundation of China (No. 12374060). Project funded by China Postdoctoral Science Foundation (No. 2023M741546, No. 2023M731530). The work at Princeton is supported by Gordon and Betty Moore Foundation (GBMF4547 and GBMF9461; M.Z.H.).

### Availability of data and materials
All data are available from the authors.

## Declarations

**Competing interests**
All authors declare that there are no competing interests.

**Author details**
$^1$Department of Physics, Southern University of Science and Technology, Shenzhen, China. $^2$Department of Physics, Princeton University, Princeton, NJ, USA. $^3$Princeton Institute for the Science and Technology of Materials, Princeton University, Princeton, NJ, USA. $^4$Materials Sciences Division, Lawrence Berkeley National Laboratory, Berkeley, CA, USA. $^5$Quantum Sciences Center, Oak Ridge, TN, USA. $^6$Quantum Science Center of Guangdong-Hong Kong-Macao Greater Bay Area (Guangdong), Shenzhen, China.

**Publisher's Note**

Springer Nature remains neutral with regard to jurisdictional claims in published maps and institutional affiliations.

**Responses to reviewers:**
Reviewer #1:
The authors propose a method combining the geometry, defect types, and interlayer relationships to distinguish different cleavage surfaces of various kagome materials, namely atomic step geometry imaging utilizing ultra-high-resolution STM. It is a highly accurate and versatile experimental method, which preventing the interlayer/intralayer distance inaccuracy induced by tunneling matrix element/surface reconstruction. On some controversial terminations with identical lattice constant and symmetry, identifying surfaces by defect types and subtle chemical-marker is effective and impressive. Furthermore, different terminations can also be confirmed by detecting distinct electronic states affected by kagome layer at negative bias voltage.

Authors: We thank the reviewer for nicely summarizing the key aspects of this perspective work. In light of the reviewer's summary of our work, we have further improved the writing of our abstract:

**Here we review scanning tunneling microscopy research on the surface determination for various types of kagome materials, including 11-type (CoSn, FeSn, FeGe), 32-type ($Fe_3Sn_2$), 13-type ($Mn_3Sn$), 135-type ($AV_3Sb_5$, A=K, Rb, Cs), 166-type ($TbMn_6Sn_6$, $YMn_6Sn_6$ and $ScV_6Sn_6$), and 322-type ($Co_3Sn_2S_2$ and $Ni_3In_2Se_2$). We first demonstrate that the measured step height between different surfaces typically deviates from the expected value of ±0.4~0.8Å, which is owing to the tunneling convolution effect with electronic states and becomes a serious issue for $Co_3Sn_2S_2$ where the expected Sn-S interlayer distance is 0.6Å. Hence, we put forward a general methodology for surface determination as atomic step geometry imaging, which is fundamental but also experimentally challenging to locate the step and to image with atomic precision. We discuss how this method can be used to resolve the surface termination puzzle in $Co_3Sn_2S_2$. This method provides a natural explanation for the existence of adatoms and vacancies, and beyond using unknown impurity states, we propose and use designer layer-selective substitutional chemical markers to confirm the validity of this method. Finally, we apply this method to determine the surface of a new kagome material $Ni_3In_2Se_2$, as a cousin of $Co_3Sn_2S_2$, and we image the underlying kagome geometry on the determined Se surface above the kagome layer, which directly visualizes the *p-d* hybridization physics. We emphasize that this general method does not rely on theory, but the determined surface identity can provide guidelines for first-principles calculations with adjustable parameters on the surface-dependent local density of states and quasi-particle interference patterns.**

This method is based on the realistic cleavage scheme in a plethora of layered kagome materials, which makes it general and powerful, not only in kagome materials but also other materials with similar lattice configurations. Therefore, I recommend potential acceptance by Quantum Frontiers. Following are some miner comments for the authors' consideration.

Authors: We thank the reviewer for the potential recommendation of our work. We address the reviewer's comments below.

1. The authors should briefly mention Fig. 3c in the main text.

Authors: In the Figure caption, we have mentioned that the right region is the K surface and the left region is the Sb surface. In the main text, we have mentioned: **Figure 3c shows a typical gradient atomic step edge, where the Sb surface is transiting to the K surface from left to right.**

2. For the termination of weak-bonding I layer, would the scanning process influence the surface condition?

Authors: For the weak-bonding I layer, the typical scanning process would not influence the surface layer. However, we agree with the reviewer that when the tip is very close to the surface (for instance with a tunneling condition V=5mV and I=5nA), the I layer (alkali layer) in $AV_3Sb_5$ would be destroyed by the tip. Some STM research groups have used this knowledge to clean out a large area of the Sb surface (by removing the above alkali islands or adatoms). Based on the reviewer's comment, we have added a note: **We also note that the**

**weakly bonded alkali islands or adatoms in 135-type of materials can be swept away by the STM tip when the tip is very close to the surface, which can help to create a large area of Sb surface for further spectroscopic research [73].**

------------------------------------------------------------------------------------------------------

Reviewer #2: In this manuscript, the authors summarized the identification of cleaved surfaces for various Kagome materials using STM. The surface determination is an important basis in the study of correlated and topological materials when using STM and ARPES, as the reconstruction, charge polarization and chemical makeup on surfaces can significantly influence the probe of intrinsic properties. In this context, I think this manuscript provides useful information on for understanding these Kagome compounds and is a good tutorial for gradate students. However, I have two concerns which hinders the recommendation of publication.

Authors: We thank the reviewer for summarizing the key aspects of this perspective. We address the reviewer's comments below.

The authors argue that this manuscript gives a "highly selfconsistent method in determining the cleavage surfaces". Personally, I do not agree with this statement. Since the intention of STM, people already know that the "height profile" in STM topography is a convolution of both surface height and electric density of states (DOS). In the past 40 years, the community have been used the same method as the authors reported here to do research with which information of topography, DOS, step edges, defects are involved. Thus, it is a bit strange to say it is "a method" that need to be reported separately. Glad to know it works for Kagome materials, but not surprising at all.

Authors: We thank the reviewer for this comment. We did not claim it is a new method. However, we would like to keep emphasizing its importance. In light of the reviewer's comment, we have removed the sentence "highly selfconsistent method in determining the cleavage surfaces". We have also emphasized that it is a challenging method in two ways in the abstract: **Hence, we put forward a general methodology for surface determination as atomic step geometry imaging, which is fundamental but also experimentally challenging to locate the step and to image with atomic precision** and in the main text: **We also note that the atomic step geometry imaging method sounds fundamental but it is also challenging to implement, requiring patience for scanning large areas and even several samples to locate the atomic step (that can take several weeks) and exceptional experimental skills for precise imaging with atomic resolution and without muti-tip effects.**

The reviewer critically mentions that "Since the intention of STM, people already know that the "height profile" in STM topography is a convolution of both surface height and electric density of states (DOS)." We agree with the reviewer that it should have been the basics in the STM community, but we see that it is also quite common for people to use this to identify surfaces (it is convenient by far from accurate). Here we give several examples cited in our work where the authors therein have used the height profile method:

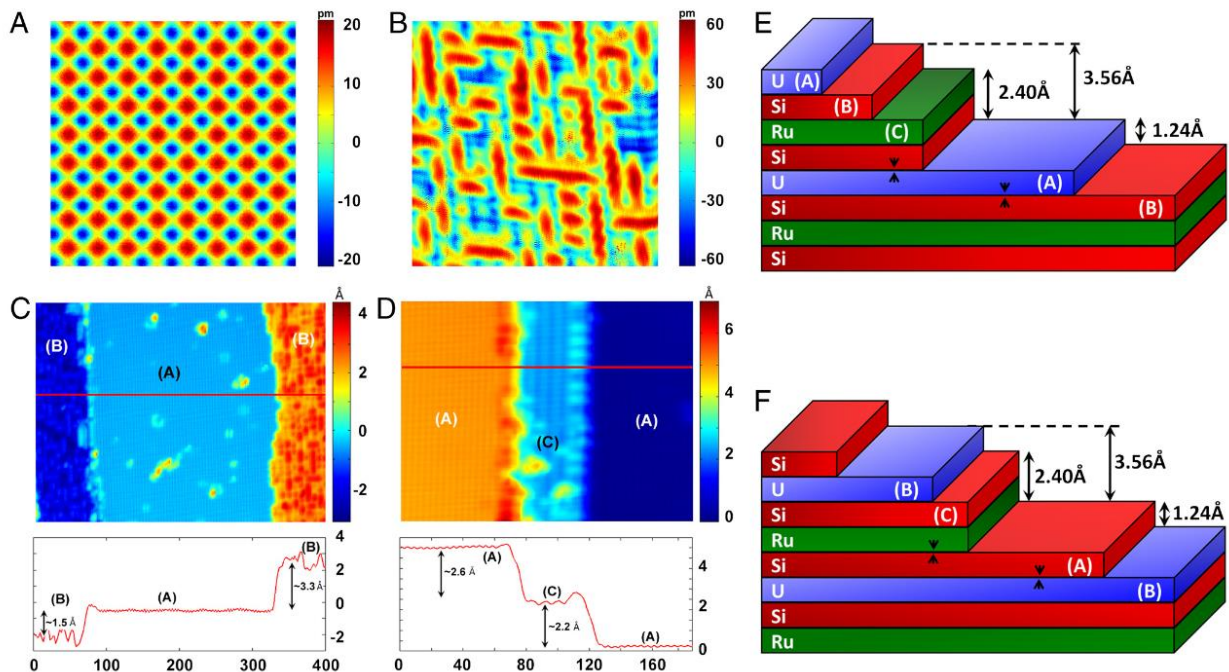

Fig. 1. STM topography. (A and B) Constant current topographic image (−200 mV, 60 pA, 33 Å) showing an atomically ordered surface (termed surface A) and (100 mV, 200 pA, 90 Å) showing an atomic layer with surface reconstruction (termed surface B), respectively. (C) The relative heights between surfaces A and B. (D) Constant current topographic image (−50 mV, 100 pA) over a 185 × 140 Å² area showing a (2 × 1) reconstructed surface (surface C) lying ~2.2 Å above surface A. A horizontal line cut through the data in C and D is shown on the bottom panels. (E) Schematic diagram illustrating the different atomic layers of $URu_2Si_2$. U is identified as the atomically ordered surface (surface A) that lies 1.24 Å above and 3.56 Å below surface B. In this case, obtaining surfaces A and B requires breaking of a single bond only (U-Si; see arrows). (F) Schematic diagram illustrating a different possibility for the cleaved surfaces, which requires the breaking of two bonds (Ru-Si and U-Si; see arrows). This scenario cannot explain why surfaces A and B occur with roughly equal probabilities. The step heights in A and B are obtained (or calculated) from ref. 6.

Data reproduced from Proceedings of the National Academy of Sciences, 2010. 107(23): p. 10383-10388. Researchers use the step height as a reference for surface determinations in $URu_2Si_2$.

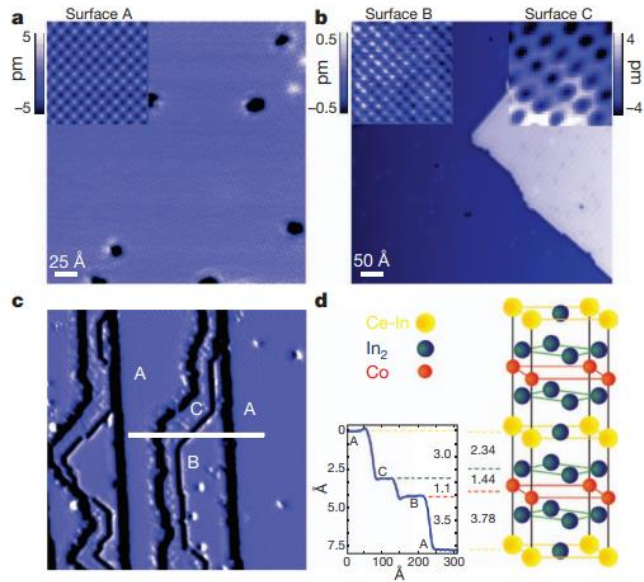

Figure 2 | STM topographies on CeCoIn$_5$. a, Constant current topographic image (+200 mV, 200 pA) showing an atomically ordered surface (termed surface A) with a lattice constant of ~4.6 Å. b, Topographic image (−200 mV, 200 pA) showing two consecutive layers: a distinct atomically ordered surface (termed surface B, lattice constant ~4.6 Å, dark blue) and a reconstructed surface (termed surface C, light blue). Insets in a and b show magnified images (45 × 45 Å$^2$) of the three different surfaces. c, Constant current topographic image (−150 mV, 365 pA) displaying all three surfaces (the derivative of the topography is shown to enhance contrast). d, A line section through the different surfaces (solid line in c) showing the relative step heights (left) compared to the bulk crystal structure (right).

Data reproduced from Nature, 2012. 486(7402): p. 201-206. Researchers use the step height as a reference for surface determinations in CeCoIn$_5$. Here we can readily see a sizeable deviation of ±0.6Å even if we assume such surface determination is correct.

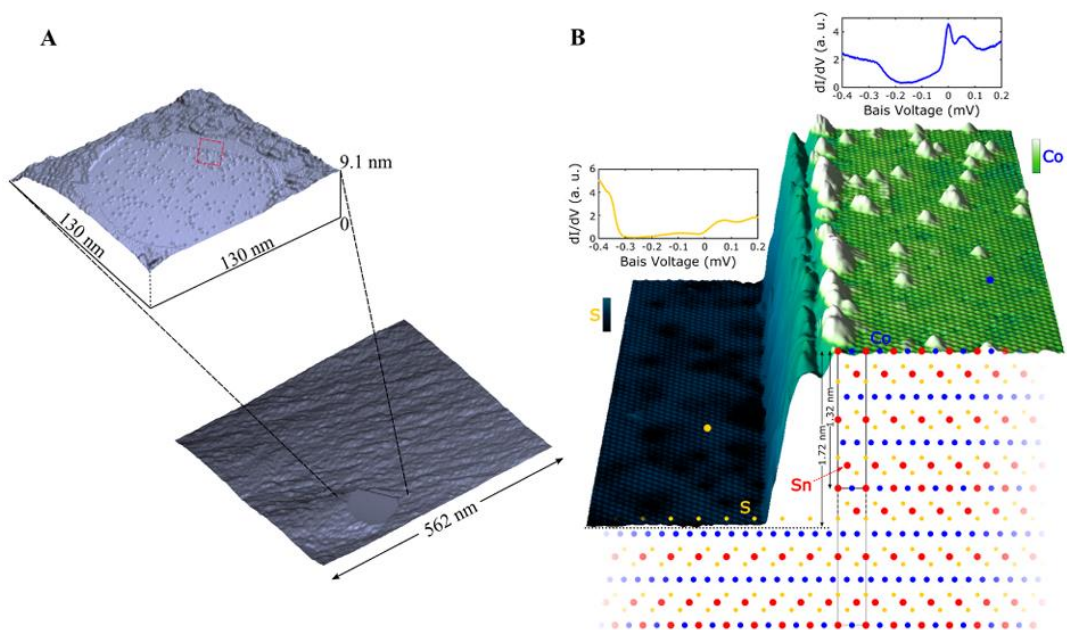

**Fig. S1.**

Co-S step height analysis. **(A)** Topography taken on a rough terrain revealing a Co terminated terrace. Inset: blown up topography of the flat Co terrace. **(B)** 3D Topographic image ($I_{set}$=150 pA, $V_{bias}$=70 meV, 32 nm X 32 nm) taken on $Co_3Sn_2S_2$ surface which includes a step between Co and S surfaces (red dashed square in A). Top surface: typical dI/dV point spectrum of the Co terminated surface measured at the blue dot. Bottom surface: typical dI/dV point spectrum of the S terminated surface, measured at the yellow dot. Superimposed crystal structure of $Co_3Sn_2S_2$ reveals the cleave planes (blue, yellow, red indicate the Co, S and Sn atoms, black solid line indicates the unit cell of the crystal).

Data reproduced from Science 365, 1286–1291 (2019). Researchers use the step height as a reference for surface determinations in $Co_3Sn_2S_2$.

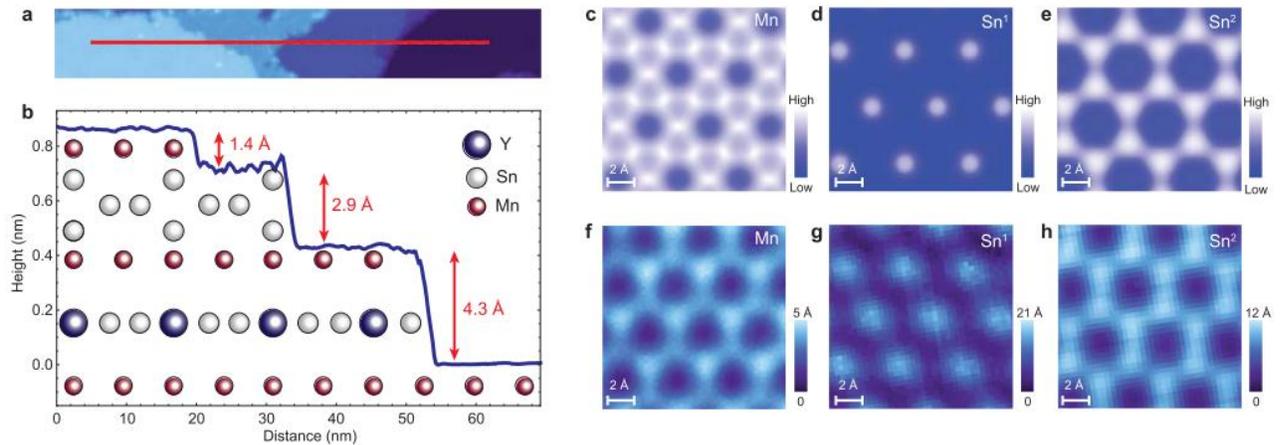

**Extended Data Fig. 2 | Surface identification based on step heights and theoretical simulations.** (**a**) STM topograph of consecutive steps. (**b**) Topographic line profile taken along the red line denoted in (a). The total height between the bottom layer and the top layer is 8.7 Å, which is a unit cell height. Based on the step heights and the nature of surface morphologies over each layer, surface terminations are identified as Mn, Sn$^1$, Mn and Mn (the tallest to the shortest terrace). (**c-e**) Theoretical simulations of STM topographs of Mn, Sn$^1$ and Sn$^2$ terminations at 30 mV bias. (**f-h**) Experimental STM topographs of Mn, Sn$^1$ and Sn$^2$ surface terminations. STM setup condition: (c-e) simulated $V_{sample} = 30$ mV. (f-h) $I_{set} = 70$ pA, $V_{sample} = 30$ mV.

Data reproduced from Nature Physics 18, 644–649 (2022). Researchers use the step height as a reference for surface determinations in YMn$_6$Sn$_6$.

It is a pity that many STM works published even in top journals still rely on the height profile method in identifying surface terminations, which is in contrast to the reviewer's comment people already know that the "height profile" in STM topography is a convolution of both surface height and electric density of states (DOS). We agree with these papers that the height profile can provide a possible hint, but it is far from decisive and could be problematic for certain systems when the error is larger than the interlayer distance. Because of the large uncertainty of this method (here it is at least ±0.8Å as shown in our Fig.2 in kagome materials), it is like gambling, while imaging the atomic step geometry does not have such issue as detailed in our paper. Therefore, we think it is important and our duty to correct this to educate the next generation of researchers.

2.For Co3Sn2S2, the surface determination might still be under debate. It is useful if the authors can discuss the difference between their assignments and those by other groups, for example, defect symmetry used by Madhavan group, QPlus AFM by HJ Gao group and Beidenkopf group. Since this manuscript does not include any new results, more discussion are needed to give more conclusive insight.

Authors: To our knowledge, several STM works use the crystals from the same group all follow the original surface determination by the Science 2019 work by Beidenkopf group mentioned above and reproduced below:

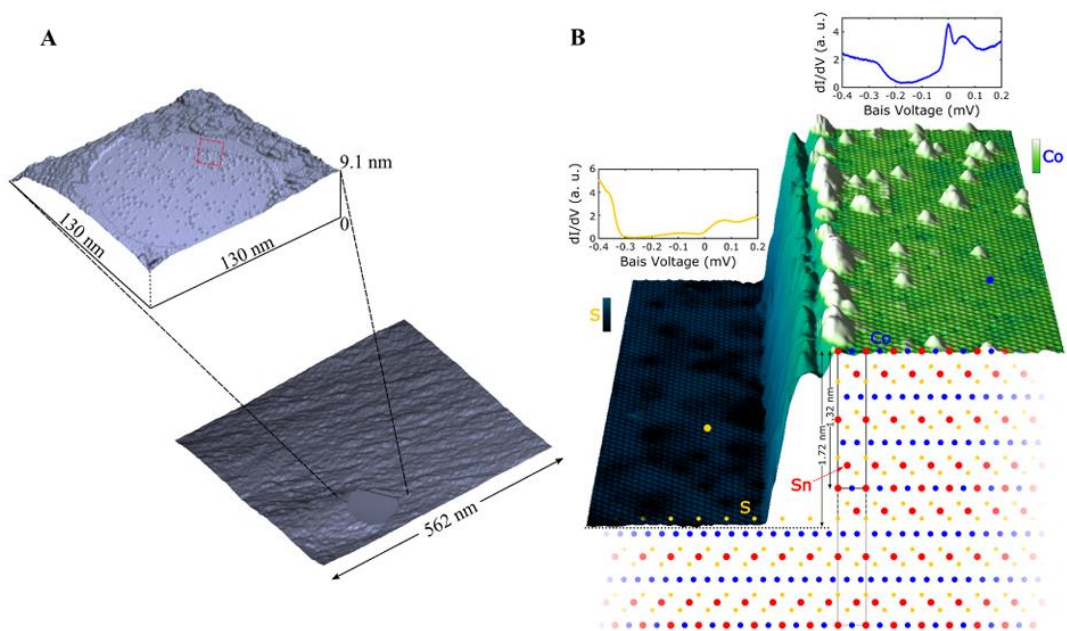

**Fig. S1.**

Co-S step height analysis. **(A)** Topography taken on a rough terrain revealing a Co terminated terrace. Inset: blown up topography of the flat Co terrace. **(B)** 3D Topographic image ($I_{set}$=150 pA, $V_{bias}$=70 meV, 32 nm X 32 nm) taken on $Co_3Sn_2S_2$ surface which includes a step between Co and S surfaces (red dashed square in A). Top surface: typical dI/dV point spectrum of the Co terminated surface measured at the blue dot. Bottom surface: typical dI/dV point spectrum of the S terminated surface, measured at the yellow dot. Superimposed crystal structure of $Co_3Sn_2S_2$ reveals the cleave planes (blue, yellow, red indicate the Co, S and Sn atoms, black solid line indicates the unit cell of the crystal).

Data reproduced from Science 365, 1286 (2019). Researchers use the step height as a reference for surface determinations in $Co_3Sn_2S_2$.

It is clear that the reviewer knows that STM topography is a convolution of both surface height and electric density of states (DOS). In light of the reviewer's comment, we have included a discussion: **We have also been aware that Ref. [23] mainly uses the step height argument for the surface identification, and it is clear from Fig. 2e, g and the basic principle of STM that the topographic height convolutes the LDOS and the geometrical corrugations that the step height argument is not decisive for termination assignment.** We have also emphasized this in the abstract: **We first demonstrate that the measured step height between different surfaces typically deviates from the expected value of ±0.4~0.8Å, which is owing to the tunneling convolution effect with electronic states and becomes a serious issue for $Co_3Sn_2S_2$ where the expected Sn-S interlayer distance is 0.6Å.**

The defect symmetry by Madhavan group uses unknown underlying defects (below the surface) with strong muti-tip effects. We first reproduce their data in surface determination below:

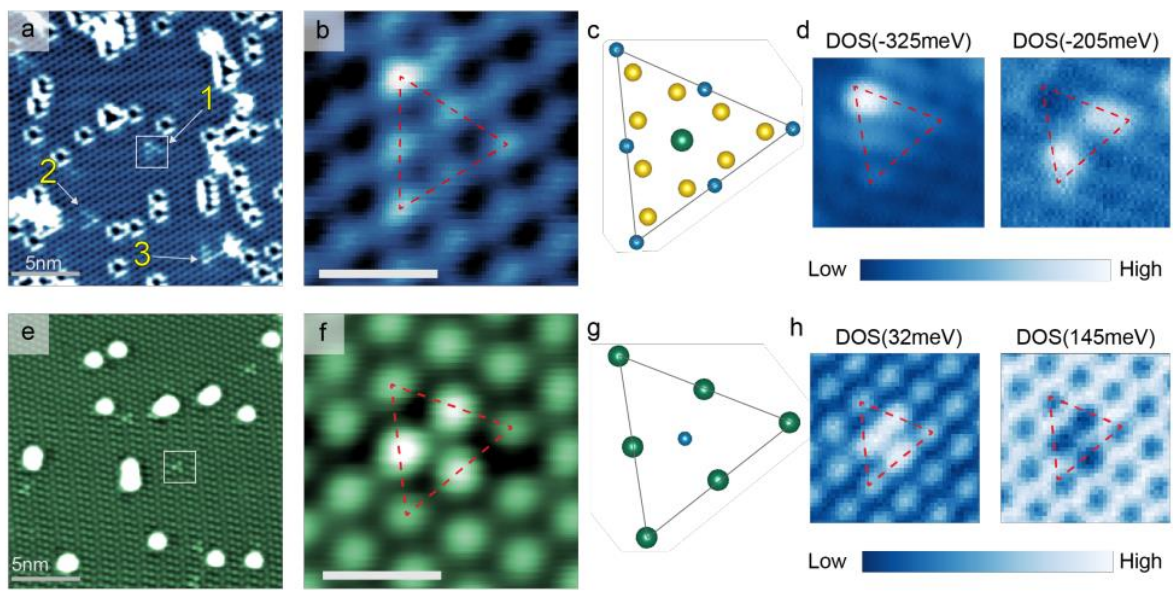

**Supplementary Fig. 2 | Topographies and corresponding crystal lattice structure of Sn and S-surfaces.**
**a,** 20 nm x 20nm topography of the surface identified from Fig 1g showing directional triangular DLBs. Three distinct defects with the bright vertex pointing up, right, and left are indicated by numbers 1, 2, and 3 respectively (I = 50 pA, V = -340 mV). **b,** Region within white box in **a** showing a zoomed in triangular DLB. The red dashed line is provided for comparison with **c**. The inset scale bar is 1nm (I = 50 pA, V = -340 mV). **c,** Schematic of the triangle in **b** as the top layer S (blue spheres) and the $Co_3Sn$ layer below. **d,** DOS maps of region in **b** at -325 meV and -205 meV showing that the density of state signature of the triangular DLB has a reduced symmetry compared to two shifted hexagonal layers, indicative of the $Co_3Sn$ layer below (I = 370 pA, V = -350 mV). **e,** 20 nm x 20 nm topography of the surface identified from Fig 1h showing clover DLBs. All DLBs found on this surface were identical. **f,** Region within white box in **e** showing a zoomed in clover DLB. The red dashed line is provided for comparison with **g**. The inset scale bar is 1nm. (I = 50 pA, V = -340 mV). **g,** Schematic of the triangle in **g** as the top layer Sn (green spheres) and the S layer below. **h,** DOS maps of region in **f** at 32 meV and 145 meV showing a trifold symmetry that is consistent with a Sn surface and a S layer below (I = 410 pA, V = -400 mV).
Data reproduced from Nat Commun 12, 4269 (2021). The defect state shape from unidentified underlying impurities is used for surface determination.

The first issue for an STM specialist is that the experimental data presented in this work is subject to the multi-tip effect: when the STM tip apex is not a single atom, the point defect will reflect the shape of the tip (the impurity atom scans the shape of the tip). This is more evidence from Fig. 1h in the same set of experiments, as reproduced below:

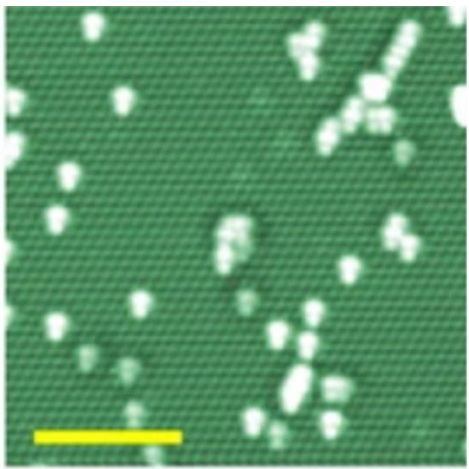
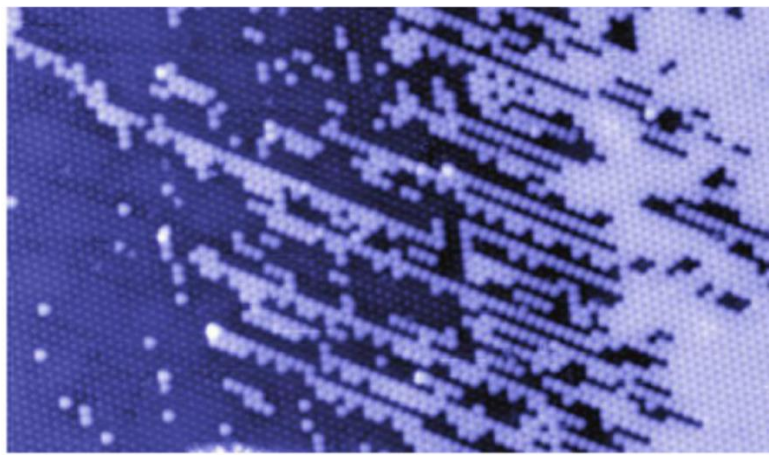

Nat Commun 12, 4269 (2021)    Nat Commun 11, 4415 (2020)

Muti-tip effect issue. In Nat Commun 12, 4269 (2021), each of the adatom impurities "scans" out the shape of the tip as made of around three atoms in an asymmetrical way. The right panel shows a similar image of the adatom surface (left side) without a noticeable multi-tip effect.

The potential multi-tip effect will enlarge the size of the impurities. The authors in Nat Commun 12, 4269 (2021) claim that the defect shape on the vacancy surface is of **six**-atom size, while on the adatom surface is of three-atom size. By contrast, in the other two STM works from the same sample source, the defect shape is always of three-atom size instead of six-atom size for the vacancy surface, as reproduced below:

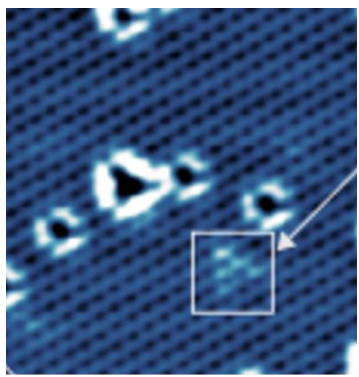
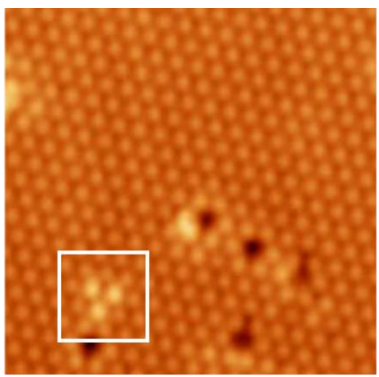
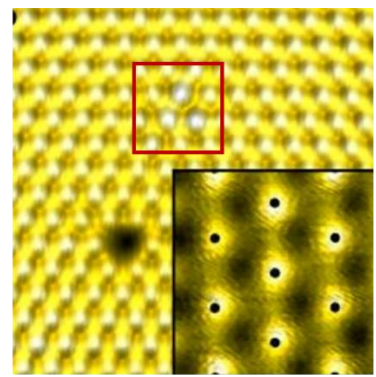

Nat Commun 12, 4269 (2021)    Nat Commun 11, 5613 (2020)    Science 365, 1286 (2019)
                                                              arXiv:1903.00509

Defect state size issue. The data from other STM works without noticeable multi-tip issues reveals impurities state size of three-atom instead of six-atom.

The potential tip effect will also lead to an unusual impurity state shape. The defect state probes the symmetry of the electronic structure, which should have less dependence on the defect type. In Nat Commun 12, 4269 (2021), the impurity states strongly break local rotational symmetry and mirror symmetry. By contrast, in another STM work with the same sample source, the vacancy defect states preserve the expected crystalline C3 symmetry and mirror symmetry, as shown below.

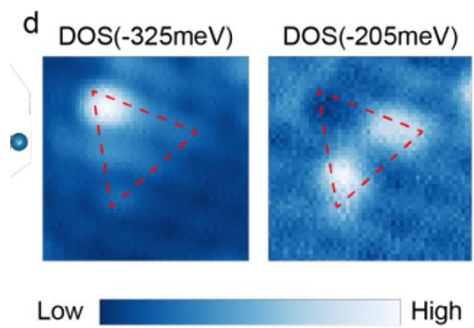
Nat Commun 12, 4269 (2021)

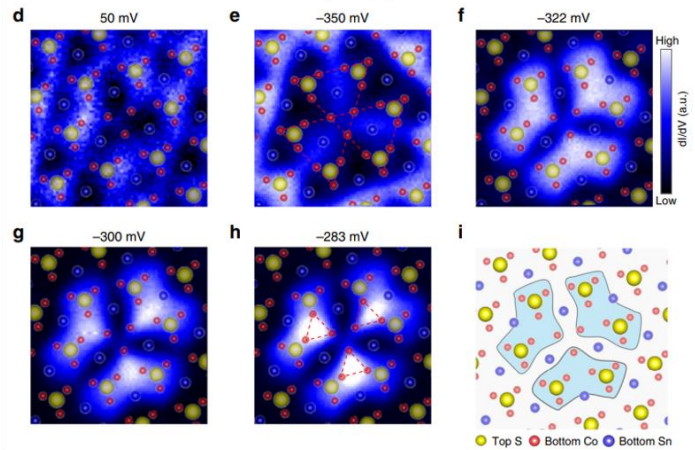
Nat Commun 11, 5613 (2020)

Defect symmetry issue. The defect states have strong asymmetry probably owing to the artificial tip effect.

In short, from a STM specialist's point of view, the scanning tip in the experiment may have systematic problems, and the defect geometry including size and shape differs substantially from other STM works on the same material without clear tip effects.

In addition to the apparent issue at the experimental data level, at the methodology level, the identity of the defect is unknown, the position of the defect is unknown, and the tunneling matrix is unknown. Even for known defects, such as Zn impurity in cuprates, its flower pattern symmetry is different from the simple theoretical model, and the tunneling matrix element effect has to be considered. Therefore, the reliability of such a methodology is largely unclear. Our response to this problem is to use a known atomic dopant as the layer-selective chemical marker. Based on the reviewer's comment, we have included more discussions:

**Following Ref. [23], the imaged impurity states shape with substantial multi-tip effects from unknown underlying (not at the surface) defects are used to provide hints for surface terminations [45]. The chemical identity, atomic position, and tunneling matrix element of the underlying defects are all unknown. Even for known underlying impurities such as Zn in cuprates [90], it is well known that the simple model would not produce the correct symmetry of the impurity states and the tunneling matrix element has to be considered. Therefore, its accuracy in identifying surface terminations with unknown underlying defects is not at the same level as the chemical marker experiment [30, 39]. We further propose using 1% Se to replace S in $Co_3Sn_2S_2$ to mark the S layer as a complementary layer-selective chemical marker experiment.**

We are not AFM experts, but in light of the reviewer's comment, we have also included a discussion on the QPlus AFM by HJ Gao group:

**Recently, the qPlus non-contact atomic force microscopy [24] has also revealed differences between the two terminations. However, the experimental data would probably not be capable of assigning the surface identifications directly and independently. Unlike the step edge imaging method, it has not been put as a general for determining surfaces in other kagome materials and other quantum materials. The first-principles calculations are then used for guiding the surface identification by matching with the atomic force microscopy data. It was our experience in 122 iron-based superconductors [30] that depending on the adjusting parameters, first principles can assign the surface terminations quite differently and oppositely. Its reliability in identifying the surfaces 122 iron-based superconductors is in doubt in the community. Even in the case of $Co_3Sn_2S_2$, first-principles calculated LDOS can always match with the experimentally observed ones with both scenarios depending on the choice of shifting the Fermi energy within ±500meV, substantially reducing its credibility. It is our understanding that in the electronic**

**structure community including angle-resolved photoemission and scanning tunneling microscopy, researchers including ourselves often seek help from first principles when the experimental data alone is inconclusive and our subjective interpretation of the data can guide the parameter tuning directions of the first-principles. Therefore, a first-principles independent methodology as demonstrated here is very crucial to firstly set up the correct and objective directions for the adjustable parameter tuning in the first-principles.** Although STM cannot directly resolve the surface atomic element, it can detect the electronic structure beneath the surface through tunneling, which provides a chance to extract the information from the underlying kagome layer. Therefore, we chose another 322-type analogue, nominating $Ni_3In_2Se_2$[91], which has the same lattice structure as $Co_3Sn_2S_2$. Figure 7a shows the side view of the step schematic model of $Ni_3In_2Se_2$, where the cleavage occurs between the In isolated layer and the Se neighboring layer. Figure 7b shows the typical step of the In and Se surface. Similar to $Co_3Sn_2S_2$, the left upper isolated layer is defined as In surface with typical vacancy defect, meanwhile, the right lower neighboring layer is defined as Se surface with In adatom defects. Although they have identical lattice symmetry, their electronic states affected by the underlying kagome layer are different. In surface is far away from the $Ni_3In$ kagome layer, so its STM topographic images show a stubborn triangular lattice for both positive (Fig. 7c) and negative bias voltage (Fig. 7d), whereas, the lower Se surface directly bonding to the $Ni_3In$ kagome layer manifests a kagome lattice feature. Hence, we observed an apparent kagome symmetry on the Se surface at the bias voltage of -200mV (Fig. 7f), whereas, an affected triangular lattice is shown in the atomic resolution image at a bias voltage of +200mV (Fig. 7e). Similar to Si(111)-7×7, lower negative bias voltage imaging facilitates the visualization of the underlying kagome layer, because their electronic states act on tunneling at this time [92, 93]. **Our observation of kagome geometry on the Se surface also directly visualized the *p-d* orbital hybridization physics as discussed in the atomic force microscopy work [24]. In short, this experimental approach and evidence provide another theory-independent way for surface identification in the 322 family.**

We have also added a note in the abstract: **We emphasize that this general method does not rely on theory, but the determined surface identity can provide guidelines for first-principles calculations with adjustable parameters on the surface-dependent local density of states and quasi-particle interference patterns.**

Finally, we note that our STM results on $Ni_3In_2Se_2$ are new. In light of the reviewer's comments, we have highlighted this in the abstract: **Finally, we apply this method to determine the surface of a new kagome material $Ni_3In_2Se_2$, as a cousin of $Co_3Sn_2S_2$, and we image the underlying kagome geometry on the determined Se surface above the kagome layer, which directly visualizes the *p-d* hybridization physics.**

We have also added more data to Fig. 7, as reproduced below:

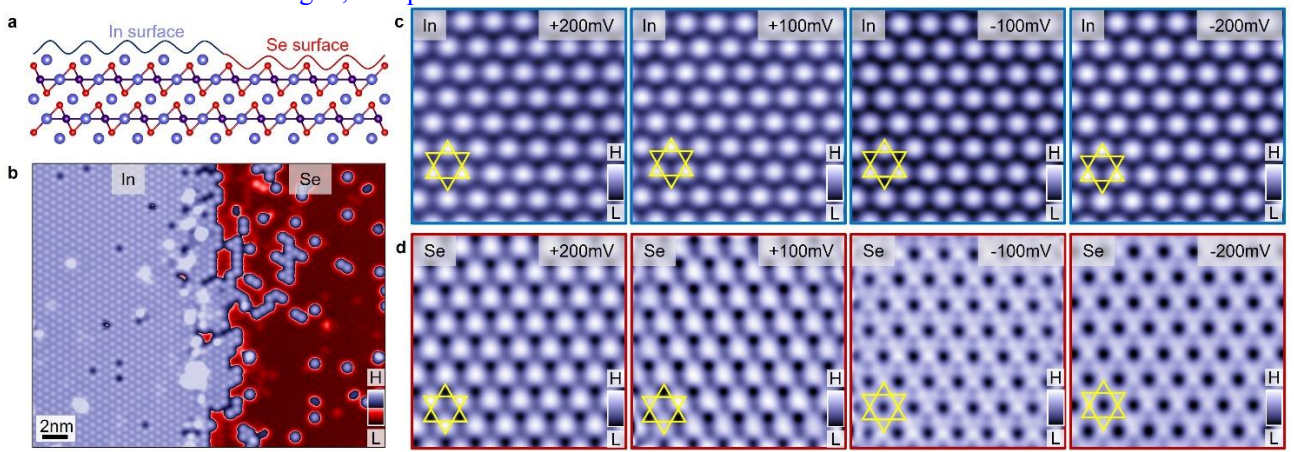

**Fig. 7 Cleavage surface identification of $Ni_3In_2Se_2$. a** Cleavage surface illustration of $Ni_3In_2Se_2$. **b** Atomically resolved topographic image of the boundary between the In surface and the Se surface (28×23 nm$^2$, $V$=100 mV, $I$=0.1 nA). **c** Topographic image of In surface obtained at a bias voltage of +200 mV, +100 mV, -100 mV and -200 mV and tunneling current of 1 nA. The yellow lines illustrate the underlying kagome lattice. **d** Topographic

image of the Se surface obtained at a bias voltage of +200 mV, +100 mV, -100 mV and -200 mV and tunneling current of 1 nA. The yellow lines illustrate the underlying kagome lattice. The data at -100 mV and -200 mV shows kagome geometry, highlighting the p-d orbital hybridization between Ni $d$-orbital and Se $p$-orbital.

For comprehensiveness, we have also included a section of 1-3 type of material $Mn_3Sn$ as Fig. 8.

Comments:

Reviewer #1: The authors have fully addressed my comments and now I'm ready to recommend for publications.

Reviewer #2: In the revised version the authors have included more data and discussions as requested. The response was in detail. I have no further questions.